\documentclass[aps,pra,twocolumn,showpacs]{revtex4-1}

\usepackage{graphicx,amsmath,amssymb}

\usepackage{bm}

\begin{document}

\title{Control and enhancement of interferometric coupling  between two photonic qubits}

\author{R. St\'{a}rek}
\affiliation{Department of Optics, Palack\' y University, 17. listopadu 1192/12,  771~46 Olomouc,  Czech Republic}

\author{M. Mi\v{c}uda}
\affiliation{Department of Optics, Palack\' y University, 17. listopadu 1192/12,  771~46 Olomouc,  Czech Republic}

\author{I. Straka}
\affiliation{Department of Optics, Palack\' y University, 17. listopadu 1192/12,  771~46 Olomouc,  Czech Republic}

\author{M. Mikov\'{a}}
\affiliation{Department of Optics, Palack\' y University, 17. listopadu 1192/12,  771~46 Olomouc,  Czech Republic}

\author{M. Je\v{z}ek}
\affiliation{Department of Optics, Palack\' y University, 17. listopadu 1192/12,  771~46 Olomouc,  Czech Republic}

\author{R. Filip}
\affiliation{Department of Optics, Palack\' y University, 17. listopadu 1192/12,  771~46 Olomouc,  Czech Republic}

\author{J. Fiur\' a\v sek}
\affiliation{Department of Optics, Palack\' y University, 17. listopadu 1192/12,  771~46 Olomouc,  Czech Republic}

\begin{abstract}
 We theoretically investigate and experimentally demonstrate a procedure for conditional control and enhancement of an interferometric coupling between two qubits 
 encoded into states of bosonic particles. Our procedure combines local coupling of one of the particles to an auxiliary mode and single-qubit quantum filtering. 
 We experimentally verify the proposed procedure using a linear optical setup where qubits are encoded into quantum states of single photons
 and coupled at a beam splitter with a fixed transmittance. With our protocol, we implement a range of different effective transmittances, 
 demonstrate both enhancement and reduction of the coupling strength, 
 and observe dependence of two-photon bunching on the effective transmittance.  To make our analysis complete, we also theoretically 
 investigate a more general scheme where each particle is coupled to a separate auxiliary mode and show that this latter 
 scheme enables to achieve higher implementation probability. We show that our approach can be extended also to other kinds of qubit-qubit interactions.
 \end{abstract}

\pacs{42.50.Ex, 03.67.Lx}

\maketitle

\section{Introduction}

The ability to design and control interactions between quantum systems represents one of the key capabilities required for quantum computing and quantum information processing \cite{Nielsen00}.
During recent years, significant theoretical and experimental effort has been devoted to development and demonstration of various elementary quantum logic gates and quantum processors 
for many physical platforms such as trapped ions \cite{Leibfried03,Blatt08,Haffner08}, Rydberg atoms \cite{Saffman10}, 
superconducting qubits \cite{Plantenberg07,DiCarlo09,Reed12, Fedorov12}, or single photons processed by linear optics  \cite{Knill01,OBrien03,Kok07}. 
While a number of important achievements have been reached, the engineering of quantum operations is in practice inevitably limited by various factors such as noise, decoherence, or limited interaction strength. 

Recently, we have addressed the issue of limited interaction strength  \cite{Micuda15} and we have shown that a weak coupling between two qubits can be conditionally 
enhanced by a combination of quantum interference and partial quantum measurement \cite{Feizpour11,Simon11} which serves as a quantum filter.
Our scheme is based on local coupling of one of the particles to an additional auxiliary internal quantum state \cite{Lanyon09}, or an auxiliary mode in case of a bosonic particle. 
In particular, we have shown that this technique allows us to conditionally implement a maximally entangling two-qubit controlled-Z gate for two qubits which are either weakly interferometrically 
coupled or whose coupling is described by a controlled-phase gate with arbitrary small conditional phase shift.

In our previous work \cite{Micuda15} we have considered an asymmetric one-sided scheme, where the coupling to an auxiliary quantum state is introduced for one of the qubits only. 
This configuration could be particularly suitable for hybrid architectures such as quantum networks combining photonic and matter qubits \cite{Kimble08}, 
where one of the quantum systems may be more difficult to control than the other. Nevertheless, it is interesting to consider also more general class of configurations
 where the coupling to an auxiliary quantum state is introduced for both particles, 
 and to fully exploit the potential of this technique to control and engineer the coupling between the qubits. 
 
 This detailed in-depth analysis is the goal of the present paper. For the sake of presentation clarity we shall mainly focus on the interferometric coupling 
 of two photons at a beam splitter with transmittance $T$. In this configuration, the transmittance provides a natural measure of the interaction strength, and the higher the transmittance the weaker the coupling. 
 For instance, implementation of a linear optical quantum CZ gate requires $T=\frac{1}{3}$ \cite{Okamoto05, Langford05,Kiesel05}. 
 Here we extend our previous analysis \cite{Micuda15} beyond the implementation of the quantum CZ gate and we investigate how to conditionally implement a coupling of two photons 
 at a beam splitter with an arbitrary effective transmittance $T_0$ when the two photons are coupled at a beam splitter with a given fixed transmittance $T$. 
 We theoretically consider both one-sided and two-sided configurations, where the coupling to an auxiliary mode is introduced for one or both photons,
respectively. We find that the two-sided scheme is generally more advantageous than the one-sided scheme, and the former yields higher implementation probability than the latter. 
Nevertheless, even with the technically simpler one-sided scheme we can achieve any $T_0\in(0,1)$ with a finite non-zero probability using any beam splitter coupling with $T\in(0,1)$.  

We experimentally demonstrate this general ability to control and tune the coupling strength with a linear optical setup 
whose core is formed by a partially polarizing beam splitter inserted inside an inherently stable interferometer formed by two calcite beam displacers. 
The qubits are encoded into states of correlated signal and idler photons generated in the process of spontaneous parametric down-conversion. 
In our experiment, $T=\frac{2}{3}$ is fixed and we demonstrate tunable effective transmittance $T_0$ which can be both higher or lower than $T$. 
We have performed full quantum process tomography of the resulting two-qubit operation and we have observed dependence of two-photon 
 Hong-Ou-Mandel interference \cite{Hong87} on the effective target transmittance $T_0$, with a clear dip close to $T_0=\frac{1}{2}$.

 For completeness, we also systematically investigate whether the more general class of two-sided schemes where the coupling to an auxiliary mode is introduced for both photons
can be exploited to increase the success rate of conditional implementation of the quantum CZ gate. Remarkably, we find that the asymmetric one-sided scheme that we have previously proposed and experimentally 
demonstrated \cite{Micuda15} is globally optimal if $T>\frac{1}{3}$ and the coupling strength needs to be increased. By contrast, if $T<\frac{1}{3}$ and the coupling strength needs to be decreased, 
then the optimal scheme is symmetric, with an equal-strength coupling to auxiliary modes introduced for both qubits. To demonstrate the general applicability of our technique, we also briefly consider a 
two-qubit interaction that results in a controlled phase gate and we show that the effective conditional phase shift can be freely tuned by our method.

 The rest of the present paper is organized as follows. In Sec. II we consider implementation of a linear optical quantum CZ gate with arbitrary 
 interferometric coupling between the two photons and we describe 
 the optimal configuration maximizing the success probability for a given $T$. In Sec. III we extend our analysis to universal control and tuning of the  
 beam splitter coupling and we determine the optimal one-sided and two-sided configurations and compare their performance.
 The experimental setup is described in Sec. IV, where we also present the experimental results. In Sec. V we discuss application of our technique to 
 interaction which results in a unitary two-qubit phase gate. Finally, brief conclusions are provided in Sec. VI.

\section{Quantum controlled-Z gate} 

The quantum controlled-Z gate \cite{Nielsen00} is a two-qubit quantum gate which introduces a $\pi$ phase shift (a sign flip), if and only if both qubits are in quantum state $|\bm{1}\rangle$,
$U_{CZ}|\bm{jk}\rangle=(-1)^{jk}|\bm{jk}\rangle$, $j,k\in \{0,1\}$.
A linear optical quantum CZ gate based on a two-photon interference on an unbalanced  beam splitter is schematically illustrated in Fig. 1 \cite{Okamoto05,Langford05,Kiesel05,Lemr11}.
 The scheme exploits the widely used dual rail encoding of qubits into quantum states of light, where each qubit is represented by a state of a single photon 
 which can propagate in two different modes.  Specifically, the logical qubit states $|\bm{0}\rangle$ and $|\bm{1}\rangle$ of photon A (B) are associated with the presence 
 of this photon in modes $A_0$ and $A_1$ ($B_0$ and $B_1$), respectively. The logical qubit states should not be confused with the Fock states and the former  
  can be expressed in terms of the latter as follows, 
 \begin{equation}
 \begin{array}{l}
 |\bm{00}\rangle_{AB}=|1010\rangle_{A_0 A_1 B_0 B_1}, \\[1mm]
 |\bm{01}\rangle_{AB}=|1001\rangle_{A_0 A_1 B_0 B_1}, \\[1mm]
 |\bm{10}\rangle_{AB}=|0110\rangle_{A_0 A_1 B_0 B_1}, \\[1mm]
 |\bm{11}\rangle_{AB}=|0101\rangle_{A_0 A_1 B_0 B_1}.
 \end{array}
 \end{equation}

\begin{figure}[!t!]
\includegraphics[width=0.98\linewidth]{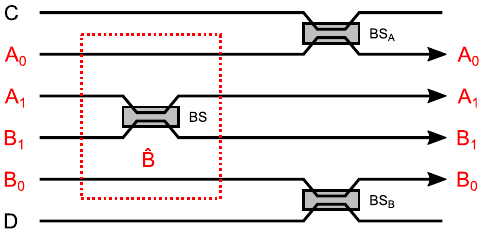}
\caption{(Color online) Linear optical quantum CZ gate operating in the coincidence basis. 
Qubit A (B) is encoded into path of a single photon propagating in a superposition of modes $A_0$ and $A_1$ ($B_0$ and $B_1$).
BS, BS$_A$ and BS$_B$ denote unbalanced beam splitters with identical intensity transmittance $T=\frac{1}{3}$. 
The dashed box indicates the conditional operation $\hat{B}$ defined in Eq. (\ref{BSstates}).}
\end{figure}

 \begin{figure*}[!t!]
\includegraphics[width=0.99\linewidth]{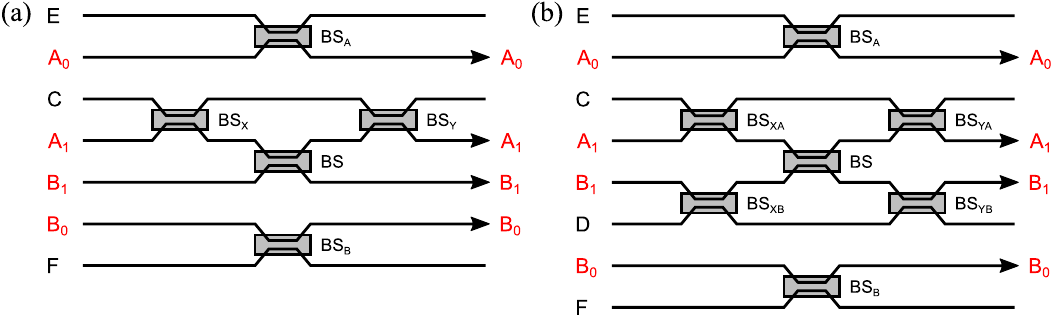}
\caption{(Color online) Quantum CZ gate with arbitrary weak interferometric coupling BS between two bosonic modes $A_1$ and $B_1$. 
(a) One-sided bypass configuration, where mode $A_1$ is coupled to the bypass mode C by beam spitter couplings BS$_X$ and BS$_Y$.
(b) Two-sided bypass configuration, where both modes $A_1$ and $B_1$ are coupled to the bypass modes C and D, respectively. 
Modes $A_0$ and $B_0$ can be attenuated by coupling to auxiliary vacuum modes E and F, respectively. }
\end{figure*}

  The gate operates in the coincidence basis \cite{Ralph02} and a successful implementation of the gate is heralded by presence of a single photon 
  in each pair of output modes $A_0$, $A_1$, and $B_0$, $B_1$. The core of the  gate consists of a two-photon interference \cite{Hong87} at an unbalanced beam splitter BS with 
transmittance $T=t^2=1/3$, which occurs only if both qubits are in logical state $|\bm{1}\rangle$. Let  $t$ and $r$ denote the amplitude transmittance and reflectance of BS, with $t^2+r^2=1$.
In the Heisenberg picture, the beam splitter coupling is described by a linear transformation of annihilation operators $\hat{a}_1$ and $\hat{b}_1$ associated with modes $A_1$ and $B_1$,
\begin{equation}
\hat{a}_{1,\mathrm{out}}=t \hat{a}_1+ r \hat{b}_1, \qquad \hat{b}_{1,\mathrm{out}}=t \hat{b}_1- r \hat{a}_1.
\end{equation}
Conditional on presence of a single photon in each pair of output modes $A_0$, $A_1$, and $B_0$, $B_1$, 
the coupling at the central beam splitter BS results in a transformation $\hat{B}$ which is diagonal in the computational basis, 
 \begin{eqnarray}
\hat{B}|\bm{00}\rangle & = & |\bm{00}\rangle, \nonumber \\ 
\hat{B}|\bm{01}\rangle & = & t|\bm{01}\rangle, \nonumber \\ 
\hat{B}|\bm{10}\rangle & = & t|\bm{10}\rangle, \nonumber \\ 
\hat{B}|\bm{11}\rangle & = & (t^2-r^2)|\bm{11}\rangle.
\label{BSstates}
\end{eqnarray}
 Note that this operation  is generally not unitary and represents a purity-preserving quantum filter. 
The auxiliary beam splitters BS$_A$ and BS$_B$ serve as additional quantum filters that attenuate the amplitudes of qubit states $|\bm{0}\rangle_A$ and $|\bm{0}\rangle_B$. 
Assuming identical transmittances of all three beam splitters BS, BS$_A$, and BS$_B$, the conditional transformation of the four basis states reads
\begin{eqnarray}
|\bm{00}\rangle &\rightarrow & T|\bm{00}\rangle, \nonumber \\ 
|\bm{01}\rangle &\rightarrow & T|\bm{01}\rangle, \nonumber \\ 
|\bm{10}\rangle &\rightarrow & T|\bm{10}\rangle, \nonumber \\ 
|\bm{11}\rangle &\rightarrow & (2T-1)|\bm{11}\rangle.
\label{CZordinary}
\end{eqnarray}
The sign flip of the amplitude of state $|\bm{11}\rangle$ required for the quantum CZ gate occurs only if $T<\frac{1}{2}$ and the value $T=\frac{1}{3}$ 
is singled out by the condition $2T-1=-T$ which ensures unitarity of the conditional gate (\ref{CZordinary}).

We will now investigate implementation of the quantum CZ gate for arbitrary interferometric coupling, i.e. for arbitrary transmittance $T$ of the central beam splitter BS 
in the optical scheme in Fig. 1. As shown in our recent work \cite{Micuda15}, this can be achieved by coupling one of the qubits to an auxiliary mode C, see Fig. 2(a). This introduces 
an additional path that allows the photon A to partly bypass the coupling with the other photon B at the central beam splitter BS \cite{Lanyon09}. 
We will first briefly review this one-sided scheme and we will then consider a more general setting where the bypass is introduced for both qubits, see Fig. 2(b).
The one-sided scheme in Fig. 2(a) may be advantageous in hybrid settings where the system B is more difficult to address and control then system A. 
Nevertheless, it is useful and instructive to analyze in depth also the more general scheme shown in Fig. 2(b), 
as it may potentially lead to higher implementation probability.

\subsection{One-sided bypass}

In the one-sided bypass scheme shown in Fig. 2(a), an auxiliary mode $C$ is introduced to partly bypass the beam splitter interaction BS \cite{Lanyon09,Micuda15}.
The beam splitters BS$_X$ and BS$_Y$ locally couple mode $A_1$ to $C$ both before and after the beam splitter interaction between modes $A_1$ and $B_1$. 
 Similarly to the standard linear optical quantum CZ gate scheme in Fig. 1, this generalized scheme also includes two beam splitters BS$_A$ and BS$_B$ that can attenuate  
  modes $A_0$ and $B_0$, respectively. In what follows, we denote by $t_j$ and $r_j$ the amplitude transmittance and reflectance of beam splitter BS$_j$.
 If we postselect on presence of a single photon in each output port of the gate then the overall transformation $\hat{W}$ implemented by the setup shown in Fig. 2(a) 
 is diagonal in the computational basis, $\hat{W}|\bm{jk}\rangle=w_{jk}|\bm{jk}\rangle$, where
 \begin{eqnarray}
 w_{00} &=  & t_A t_B , \nonumber \\
 w_{01} & = &t_A t, \nonumber \\
w_{10} & = &\left(t t_X t_Y-r_Xr_Y\right)t_B, \nonumber \\
w_{11} & = &\left(2t^2-1\right)t_X t_Y-tr_Xr_Y.
\end{eqnarray}
 
The quantum CZ gate is conditionally implemented provided that $w_{00}=w_{10}=w_{01}=-w_{11}$, which yields the conditions
 $t_A=tt_Xt_Y-r_Xr_Y$, $t_B=t$,  and 
 \begin{equation}
 \frac{r_Xr_Y}{t_Xt_Y}=\frac{3T-1}{2t}.
 \label{XYconditionCZ}
 \end{equation}
 The probability of implementation of the gate can be expressed as  $P_I=|t_A t_B|^2$ and after some algebra we obtain 
 \begin{equation}
 P_I=\frac{1}{4} (1-T)^2 t_X^2t_Y^2.
 \end{equation}
 Formula (\ref{XYconditionCZ}) describes a one-parametric class of schemes implementing the quantum CZ gate. Using Eq. (\ref{XYconditionCZ}) 
 we can express $t_Y^2$ in terms of $t_X^2$, insert the resulting expression into formula for $P_I$, and search for its maximum over $t_X$. 
 This optimization can be performed analytically  and we find that the implementation probability is maximized  by a symmetric configuration, where
 \begin{equation}
 t_X^2=t_Y^2=\frac{2t}{2t+|1-3T|}.
 \end{equation}
 For this choice of coupling to mode C we get 
 \begin{equation}
 P_I=\frac{(1-T)^2T}{(2t+|1-3T|)^2}.
 \label{PSCZ}
 \end{equation}
The quantum interference conditionally enhances the coupling of modes $A_1$ and $B_1$ although this interaction is
 partially bypassed by coupling mode $A_1$ with mode $C$. 
 
 \subsection{Two-sided bypass}
 We will now turn our attention to the more general class of schemes with  coupling to auxiliary modes introduced for both qubits, c.f. Fig. 2(b). 
 We can see that the coupling of modes $A_1$ and $B_1$ to auxiliary modes C and D, respectively, 
 is provided by four beam splitters $BS_{XA}$, $BS_{YA}$, $BS_{XB}$, and $BS_{YB}$. The conditional transformation $\hat{W}$ introduced above 
 remains diagonal in the computational basis even for this extended scheme, only the expressions for the amplitudes $w_{jk}$ become more involved,
 \begin{eqnarray}
 w_{00}&=& t_At_B, \nonumber \\
 w_{01}&=& t_A (t_{XB}t_{YB}t-r_{XB}r_{YB}), \nonumber \\
 w_{10}&=& t_B (t_{XA}t_{YA}t-r_{XA}r_{YA}), \nonumber \\
 w_{11}&=& w_{01}w_{10}w_{00}^{-1}-t_{XA}t_{XB}t_{YA}t_{YB}r^2.
 \label{wjkgeneral}
 \end{eqnarray}
 The quantum CZ gate is implemented provided that
 \begin{eqnarray}
 t_A =t_{XA}t_{YA}t-r_{XA}r_{YA},  \nonumber \\
 t_B = t_{XB}t_{YB}t-r_{XB}r_{YB},
 \label{tAtBgeneralCZ}
 \end{eqnarray}
 and
 \begin{equation}
 \frac{r_{XA}r_{YA}}{t_{XA}t_{YA}}=t-\frac{1}{2}\frac{t_{XB}t_{YB}(1-T)}{t_{XB}t_{YB}t-r_{XB}r_{YB}}.
 \label{XYconditionCZgeneral}
 \end{equation}
 This formula generalizes Eq. (\ref{XYconditionCZ}) and describes a three-parametric class of schemes implementing the quantum CZ gate with probability $P_I=t_A^2t_B^2$.
 
 \begin{figure}[t]
\centerline{\includegraphics[width=\linewidth]{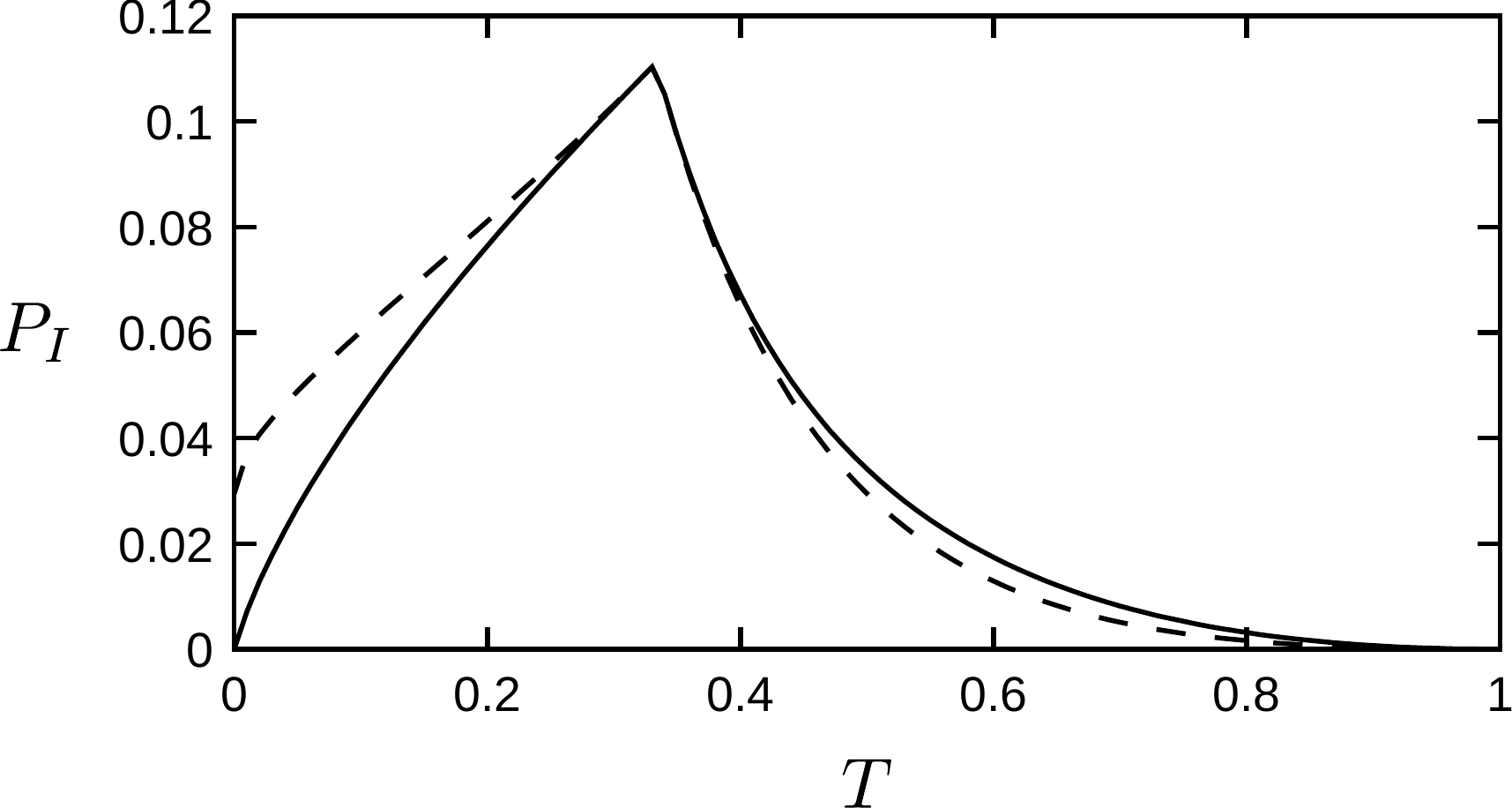}} 
\caption{Probability of implementation $P_I$ of a linear optical quantum CZ gate with the optimal one-sided bypass configuration (solid line) and with the symmetric two-sided bypass configuration (dashed line) 
are plotted in dependence on transmittance $T$ of beam splitter BS.}
 \end{figure}
 
 As an important special case, let us investigate a symmetric configuration, where the coupling to the auxiliary mode is the same for both qubits, 
 \begin{equation}
 \begin{array}{c}
  t_{XA}=t_{XB}, \qquad t_{YA}=t_{YB}, \\[1mm]
  r_{XA}=r_{XB}, \qquad r_{YA}=r_{YB}.
 \end{array}
 \end{equation}
 In this case, Eqs. (\ref{tAtBgeneralCZ}) and (\ref{XYconditionCZgeneral}) yield
 \begin{equation}
 \frac{r_{XA}r_{YA}}{t_{XA}t_{YA}}=t \pm \frac{r}{\sqrt{2}},
 \end{equation}
 and
 \begin{equation}
 t_A=t_B = \pm \frac{r}{\sqrt{2}}t_{XA} t_{YA}.
  \end{equation}
The probability of implementation of the CZ gate is maximized when $t_{XA}^2=t_{YA}^2=\sqrt{2}/(\sqrt{2}+|r-\sqrt{2}t|)$, and we get
\begin{eqnarray}
\tilde{P}_I=\frac{(1-T)^2}{(\sqrt{2}+|r-\sqrt{2}t|)^4}.
\label{PStildeCZ}
\end{eqnarray}

The implementation probability $P_I$ of the optimal one-sided scheme and $\tilde{P}_I$ of the symmetric two-sided scheme are plotted Fig.~3. We can see that $P_I>\tilde{P}_I$ when
$T>\frac{1}{3}$ while  $P_I<\tilde{P}_I$ when $T<\frac{1}{3}$. We have performed numerical optimization of the implementation probability over the full three-parametric class (\ref{XYconditionCZgeneral}) 
of the general two-sided schemes. This calculation reveals that for $T>\frac{1}{3}$ the maximum achievable implementation probability is equal to the probability $P_I$ given by Eq.~(\ref{PSCZ}), hence the  
one-sided scheme is globally optimal in this regime. On the other hand, if  $T<\frac{1}{3}$ then the symmetric two-sided scheme is globally optimal and $\tilde{P}_{I}$ is the maximum achievable implementation probability.

\section{Universal control of interferometric coupling}
The realization of the linear optical CZ gate with arbitrary interferometric coupling investigated in the previous Section can be seen as an implementation of 
a beam splitter with effective transmittance $T_0=\frac{1}{3}$ using a beam splitter with a different transmittance $T$. In this Section we extend our study beyond the 
quantum CZ gate and we consider implementation of a beam splitter with arbitrary transmittance $T_0$. Specifically, we will investigate conditional implementation of the two-qubit transformation
$\hat{B}$ using the schemes with one-sided and two-sided bypass as depicted in Fig.~2.

Conditional on presence of a single photon in each pair of output modes $A_0$, $A_1$ and $B_0$, $B_1$, the interferometric schemes in Fig. 2 
implement the beam splitter coupling (\ref{BSstates}) with transmittance $T_0$ provided that
\begin{equation}
\frac{w_{01}}{w_{00}}=\frac{w_{10}}{w_{00}}=t_0, \qquad \frac{w_{11}}{w_{00}}=t_0^2-r_0^2,
\label{BSWconditions}
\end{equation}
where the coefficients $w_{jk}$ are given by Eq. (\ref{wjkgeneral}). The first two of these conditions yield the following expressions for transmittances $t_A$ and $t_B$,
\begin{eqnarray}
 t_A =\frac{1}{t_0}\left(t_{XA}t_{YA}t-r_{XA}r_{YA}\right),  \nonumber \\
 t_B = \frac{1}{t_0}\left(t_{XB}t_{YB}t-r_{XB}r_{YB}\right).
 \label{tAtBgeneralBS}
 \end{eqnarray}
 Note that for certain parameter values it may happen  that $|t_A|>1$. In such case one should attenuate mode $A_1$ by factor of $1/t_A$ while the mode $A_0$ is not attenuated at all. 
 Similarly, if  $|t_B|>1$ then mode $B_1$ should be attenuated by $1/t_B$. 
 The final condition in Eq. (\ref{BSWconditions}) provides the following relation between the parameters of the four beam splitters that implement the two bypasses,
 \begin{equation}
 \frac{r_{XA}r_{YA}}{t_{XA}t_{YA}}=t-\frac{t_0^2}{r_0^2}\frac{t_{XB}t_{YB}r^2}{t_{XB}t_{YB}t-r_{XB}r_{YB}}.
 \label{XYconditionBSgeneral}
 \end{equation}
 Choice of beam splitter parameters satisfying Eq. (\ref{BSWconditions}) ensures that the implemented transformation reads $\hat{W}=\sqrt{P_I}\hat{B}$, where $P_I$ is the implementation probability.
 If $|t_A|\leq 1$ and   $|t_B|\leq 1$ then
 \begin{equation}
 P_I=t_A^2t_B^2.
 \end{equation}
 If $|t_A|>1$ according to Eq. (\ref{tAtBgeneralBS}) then no attenuation is applied to mode $A_0$ and $P_I=t_B^2$. Similarly, if $|t_B|>1$ then $P_I=t_A^2$.
 Recall that the transformation $\hat{B}$ is itself non-unitary, hence only probabilistic. The implementation probability $P_I$ of our procedure represents an additional factor, 
 which further reduces the success probability of the resulting transformation $\hat{W}$.
 
 We have performed numerical optimization of  $P_I$ over the whole three-parametric class of configurations specified by Eqs. (\ref{tAtBgeneralBS}) and (\ref{XYconditionBSgeneral}). 
 Based on this numerical analysis we have identified optimal configurations maximizing $P_I$. For $T_0>T$, the optimal configuration is fully symmetric, with 
 \begin{equation}
 \begin{array}{c}
 t_{XA}=t_{XB}=t_{YA}=t_{YB},  \\[1mm]
 r_{XA}=r_{XB}=-r_{YA}=-r_{YB},  \\[1mm] 
 \qquad  t_A=t_B.
 \end{array}
 \end{equation}
 Using Eqs. (\ref{tAtBgeneralBS}) and (\ref{XYconditionBSgeneral}) we obtain
 \begin{equation}
 t_{XA}^2=\frac{r_0}{t_0r+r_0(1-t)},\qquad t_A=\frac{r}{t_0r+r_0(1-t)}.
 \end{equation}
It can be shown analytically that $t_{XA}<1$ and $t_A<1$ provided that $t_0>t$. The implementation probability in this case thus reads 
\begin{equation}
\tilde{P}_I^{-}=\left( \frac{r}{t_0r+r_0(1-t)}\right)^4.
\end{equation}
 If $T_0<T$, then the optimal configuration is specified by conditions
\begin{equation}
t_{XA}=t_{YA}, \qquad t_{XB}=t_{YB}, \qquad t_B=1,
\end{equation}
and $r_{YA}=r_{XA}$, $r_{YB}=r_{XB}$. After some algebra, we get 
\begin{equation}
t_{XA}^2=\frac{(1+t)r_0^2}{(1+t)^2r_0^2-(1+t_0)r^2t_0}, \qquad t_{XB}^2=\frac{1+t_0}{1+t}, 
\end{equation}
and
\begin{equation}
\tilde{P}_I^{+}=t_A^2=\left(\frac{r^2}{(1+t)^2(1-t_0)-r^2 t_0}\right)^2.
\end{equation}

 \begin{figure}[t]
\centerline{\includegraphics[width=\linewidth]{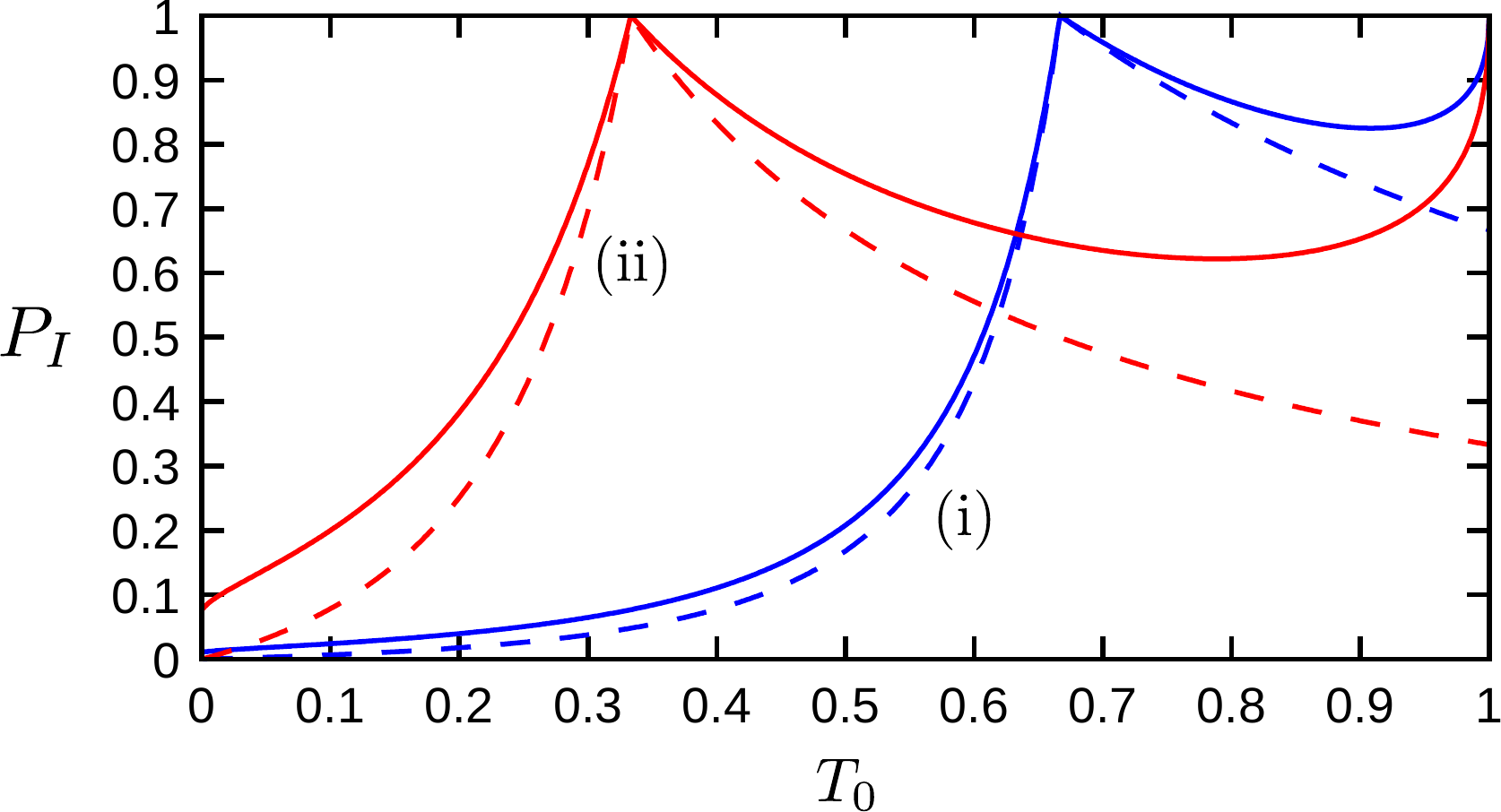}} 
\caption{(Color online) Probability of implementation $P_I$ of beam splitter coupling with transmittance $T_0$ is plotted for two values 
of the actual beam splitter transmittance: (i)  $T=\frac{2}{3}$ (blue lines) and (ii) $T=\frac{1}{3}$ (red lines). Shown is the implementation probability for the optimal 
two-sided bypass configuration (solid lines) as well as for the optimal one-sided bypass configuration (dashed lines).}
 \end{figure}

Let us now consider implementation of an arbitrary beam splitter coupling with the one-sided bypass scheme shown in Fig. 2(a). Since $t_{XB}=t_{YB}=1$ in this case, we have
\begin{equation}
t_B=\frac{t}{t_0}.
\end{equation}
If $T_0<T$, then $t_B>1$, hence mode $B_1$ has to be attenuated by factor of $t_0/t$, and the optimal configuration maximizing the implementation  probability $P_I$ is symmetric,
with 
\begin{equation}
t_{XA}=t_{YA}=\sqrt{\frac{t r_0^2}{t r_0^2+t^2-t_0^2 }},
\end{equation}
and $r_{XA}=r_{YA}$. On inserting these expressions into the formula (\ref{tAtBgeneralBS}) for $t_A$, we get
\begin{equation}
P_{I}^{+}=t_A^2=\left( \frac{t_0 r^2}{t r_0^2+t^2-t_0^2} \right)^2.
\label{PIplusBS}
\end{equation}
If $T_0>T$, then $t_B<1$ and one can choose the amplitude transmittances $t_{XA}$ and $t_{YA}$ such that $t_A=1$. This is achieved for 
\begin{eqnarray}
t_{XA}^2&=&\frac{1}{2}\left(1+x-y+\sqrt{(1+x-y)^2-4x}\right), \nonumber \\
t_{YA}^2&=&\frac{1}{2}\left(1+x-y-\sqrt{(1+x-y)^2-4x}\right),
\end{eqnarray}
where
\begin{equation}
 x=\frac{t^2 r_0^4}{t_0^2 r^4}, \qquad
 y=\left(t_0-\frac{t^2 r_0^2}{t_0 r^2}\right )^2.
\end{equation}
Since $t_A=1$, the success probability of this optimal configuration with one-sided bypass reads $t_B^2$, hence
\begin{equation}
P_I^{-}=\frac{T}{T_0}.
\label{PIminusBS}
\end{equation}
The implementation probability $\tilde{P}_I$ achieved by the optimal two-sided bypass scheme as well as implementation probability $P_I$ achieved by the optimal scheme with one-sided bypass 
are plotted in Fig.~4 for two different values of $T$.
We can see that the two-sided scheme generally outperforms the one-sided scheme. In particular, a totally reflecting beam splitter with $T_0=0$ 
can be implemented with a non-zero probability $\tilde{P}_I=r^4/(1+t)^4$ with the two-sided bypass scheme, while this probability is equal to zero for the one-sided scheme.
Also, $\tilde{P}_I=1$ in the limit $T_0=1$, because with the two-sided bypass it is trivial to deterministically switch off the qubit coupling at the beam splitter BS. 
By contrast, with a single-sided bypass such switching off of the coupling can be performed only probabilistically, and the corresponding implementation probability reads $P_I=T$.

\section{Experiment}

We have experimentally tested the control of interferometric coupling between two photonic qubits with the experimental setup depicted in Fig. 5. 
Time-correlated orthogonally polarized photon pairs were generated in the process of frequency-degenerate collinear type II spontaneous parametric downconversion in a nonlinear crystal pumped by a laser diode 
with central wavelength of $405$~nm (not shown in Fig.~5). The pump beam was removed by a dichroic mirror and the downconverted signal and idler photons at $810$~nm 
were spatially separated on a polarizing beam splitter, coupled into single-mode fibers and guided to the two input ports of the 
bulk interferometer shown in Fig.~5, where they were released into free space. Qubit A was encoded into path of the signal photon propagating in an inherently stable 
Mach-Zehnder interferometer formed by two calcite beam displacers BD \cite{Lanyon09,Micuda13,Micuda15}. 
Specifically, the qubit state $|\bm{0}\rangle_A$ corresponds to the horizontally polarized photon  propagating in the upper interferometer arm, while the state $|\bm{1}\rangle_A$ is represented by a vertically polarized photon 
propagating in the lower interferometer arm. Preparation of an arbitrary input state of qubit A is achieved by combination of half-wave plate HWP and quarter-wave plate QWP followed by beam displacer BD$_1$, 
which introduces a transversal spatial offset between the vertically and horizontally polarized beams and converts the polarization encoding into path encoding. 
The second calcite beam displacer BD2 converts the path encoding back into polarization which ensures that the output state
of qubit A can be analyzed with the use of a standard single-photon polarization detection block DB which consists of a HWP, QWP, polarizing beam splitter PBS, and single-photon detectors. 
Qubit B is encoded into polarization state of the idler qubit and the computational states $|\bm{0}\rangle_B$ and  $|\bm{1}\rangle_B$ are represented by horizontal and vertical polarization of the idler photon, respectively.

\begin{figure}[!t!]
\centerline{\includegraphics[width=\linewidth]{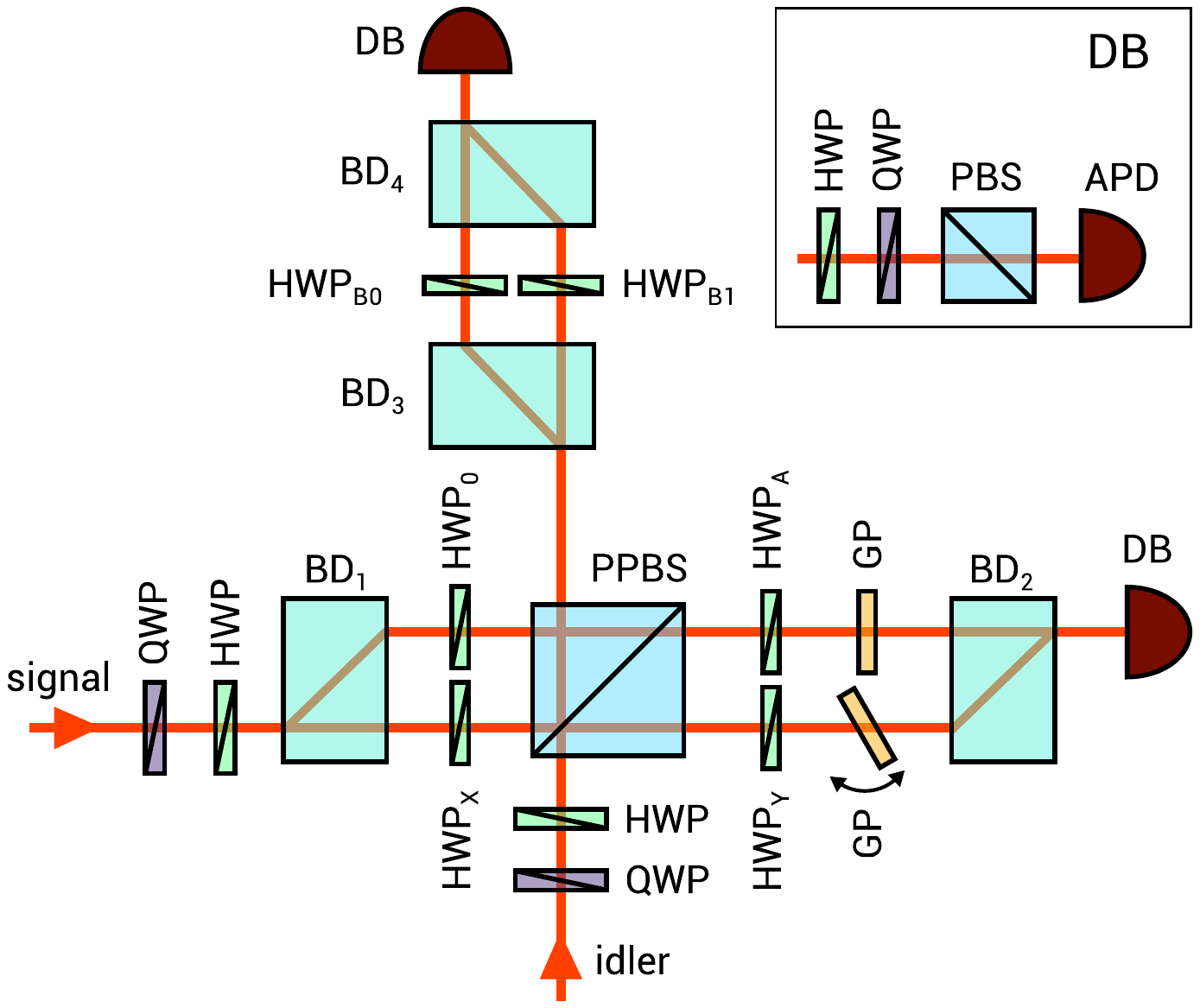}}
\caption{(Color online) Experimental setup. HWP - half-wave plate, QWP - quarter-wave plate, 
PPBS - partially polarizing beam splitter with transmittances $T_V=\frac{2}{3}$ and $T_H=1$ for vertical and horizontal polarizations, respectively,
PBS - polarizing beam splitter, BD - calcite beam displacer. 
The inset shows the single-photon polarization detection block DB which consists of a HWP, QWP, PBS and single-photon detectors APD.}

\end{figure}

\begin{figure*}[!t!]
\centerline{\includegraphics[width=\linewidth]{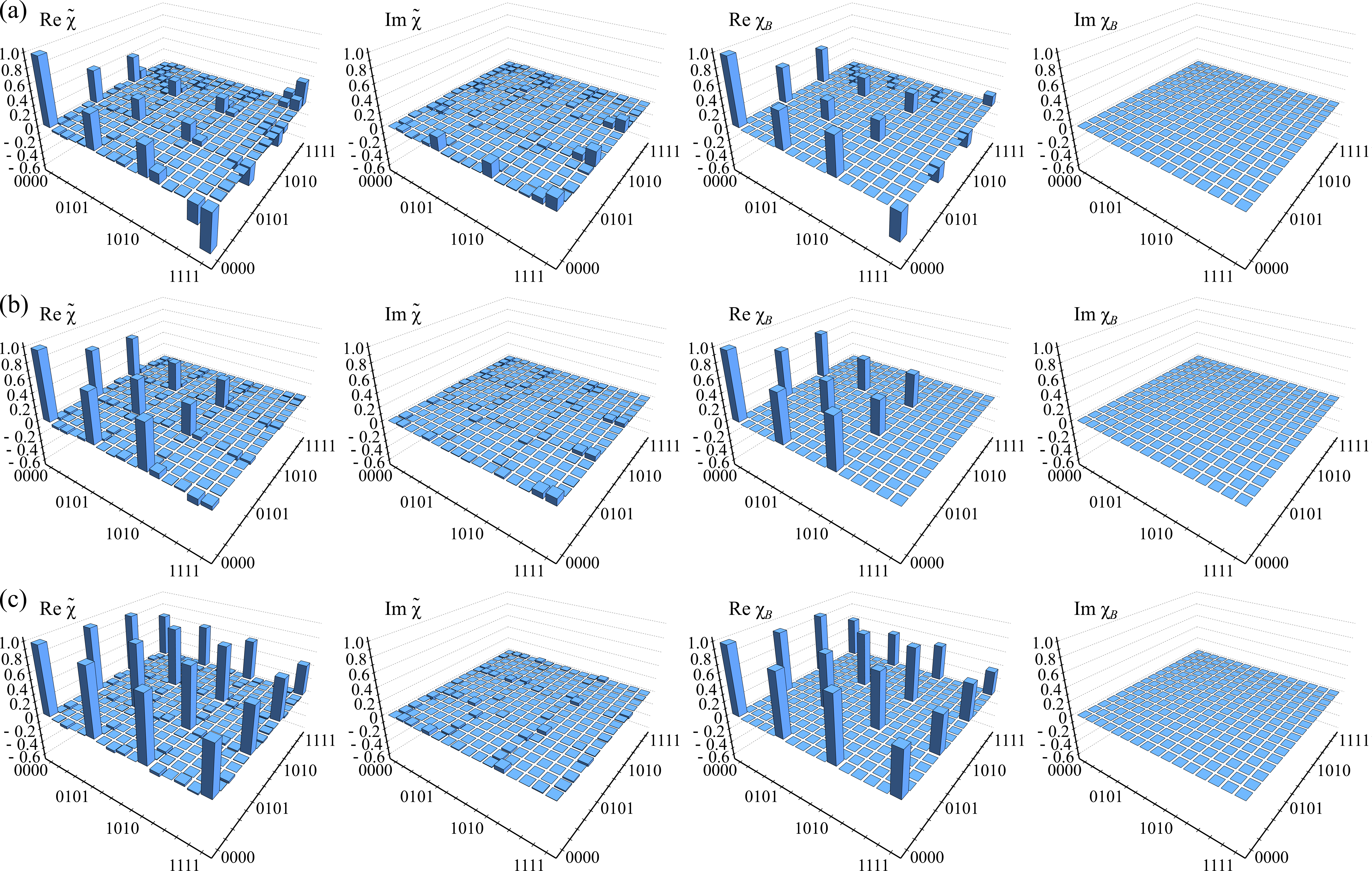}}
\caption{(Color online) Quantum process matrices of conditional two-qubit operations induced by interferometric coupling. Results are shown for three different values 
of target beam-splitter transmittance $T_0=0.3$ (a), $T_0=0.5$ (b), and $T_0=0.8$ (c). The nominal transmittance of the beam splitter reads $T=2/3$. 
The first two columns contain real and imaginary parts of process matrices $\tilde{\chi}$ reconstructed from the experimental data, 
while the third and fourth column display real and imaginary parts of the theoretical matrices $\chi_B$ for comparison.}
\label{fig:choi}
\end{figure*}

The interferometric coupling between the two photons is provided by a partially polarizing beam splitter PPBS, which is fully transparent for horizontally polarized beams, $T_H=1$, 
while it partially reflects vertically polarized beams, $T_V=\frac{2}{3}$. The interference at PPBS thus corresponds to the interference of modes $A_1$ and $B_1$ at a beam splitter BS in Fig. 2(a), with $T=T_V=\frac{2}{3}$.
A controllable attenuation of mode $A_0$ is accomplished by a half-wave plate HWP$_A$, whose rotation angle $\phi_A$ determines the attenuation factor, $t_A=\sin(2\phi_A)$. The rotated wave plate transforms 
initially horizontally polarized beam onto a linearly polarized beam and only the vertically polarized component is transmitted to the output port through the calcite beam displacer BD$_2$. The bypass on qubit A 
is implemented by exploiting the polarization degree of freedom and the auxiliary mode C is represented by a horizontally polarized mode in the lower interferometer arm. Coupling between mode $A_1$ and the auxiliary mode $C$ 
is provided by two half-wave plates HWP$_X$ and HWP$_Y$ that play the role of beam splitters BS$_X$ and BS$_Y$ in Fig. 2(a). 
To balance the paths, we make the interferometer symmetric with identical wave plates and glass plates inserted into both of its arms.
By tilting one of the glass plates GP we can control the relative phase shift between the two interferometer arms and set it to zero.

Our protocol also requires tunable attenuation of output modes $B_0$ and $B_1$. We accomplish this with the use of another interferometer formed by a pair of calcite beam displacers 
BD$_3$ and BD$_4$with a HWP inserted in each arm of the interferometer. 
This configuration allows us to selectively attenuate mode $B_0$ or $B_1$ by suitable rotations of either HWP$_{B0}$ or HWP$_{B1}$, respectively. 
After filtering, polarization state of qubit B is analyzed with the help of a second polarization detection block DB. 
The scheme operates in the coincidence basis and its successful operation is indicated by coincidence detection of two photons, one by each detection block DB. 
For additional details about the experimental setup, see Refs. \cite{Micuda13, Micuda15}.

Following the optimal one-sided bypass protocol theoretically described in Sec. III, we have used our setup to implement a beam splitter coupling $\hat{B}$ 
with 11 different effective transmittances $T_0=0.3+0.05j$, where $j=0,1,\ldots,10$. 
We have performed full quantum process tomography of the implemented two-qubit operations. 
We have probed the operation with $36$  different product two-qubit states, where each of the qubit is chosen to be in one of the six states 
$|\bm{0}\rangle$, $|\bm{1}\rangle$, $\frac{1}{\sqrt{2}}(|\bm{0}\rangle\pm|\bm{1}\rangle)$, or $\frac{1}{\sqrt{2}}(|\bm{0}\rangle\pm i|\bm{1}\rangle)$.
We label these $36$ two-qubit input states by an integer $m$. For each input state $m$, the number of coincidence detections $C_{mn}$  corresponding to projection of the output photons onto a two-qubit product state $n$  
was measured for a fixed time interval of $1$~s. Utilizing the Choi-Jamiolkowski isomorphism \cite{Jamiolkowski72,Choi75}, we can represent a two-qubit quantum operation 
 by a positive semidefinite operator $\chi$ acting on Hilbert space of four qubits (two input qubits and two output qubits). 
 The quantum process matrix $\chi$ was reconstructed from the measured coincidences $C_{mn}$ using the maximum likelihood estimation procedure \cite{Paris04}.

The process matrix $\chi_{B}$ representing the target operation (\ref{BSstates}) is proportional to a density matrix of a pure entangled four-qubit state, $\chi_{B}=|\Phi_{B}\rangle\langle \Phi_{B}|$,
where $|\Phi_B\rangle$ is obtained by applying the operation $\hat{B}$ to one part of a maximally entangled four-qubit state,
\begin{equation}
|\Phi_B\rangle=\hat{I} \otimes \hat{B} \sum_{j,k=0}^1 |\bm{jk}\rangle |\bm{jk}\rangle.
\end{equation}
Explicitly, we have
\begin{equation}
|\Phi_{B}\rangle= |\bm{0000}\rangle+t|\bm{0101}\rangle+t|\bm{1010}\rangle+(t^2-r^2)|\bm{1111}\rangle.
\end{equation}
For ease of visual comparison between theory and experiment, we introduce normalized process matrices 
\begin{equation}
\tilde{\chi}= \frac{\chi}{\langle \bm{0000}|\chi|\bm{0000}\rangle},
\end{equation}
This normalization ensures that   $\langle \bm{0000}|\tilde{\chi}|\bm{0000}\rangle=1$ which holds for the matrices $\chi_{B}$ of the ideal target operations $\hat{B}$. 
In Fig.~6 we plot the reconstructed quantum process matrices $\tilde{\chi}$ for three target transmittances $T_0=0.3$, $T_0=0.5$, and $T_0=0.8$, 
together with the corresponding ideal process matrices $\chi_{B}$. Since $T=\frac{2}{3}$, the cases $T_0=0.3$ and $T_0=0.5$ 
correspond to enhancement of the interferometric coupling, while the case $T_0=0.8$ illustrates reduction of the coupling strength.  
 We observe a good agreement between the experimental and theoretical process matrices.  Since $\chi_{B}$ is proportional to a density matrix of a pure state, 
 we can quantify the similarity between $\chi$ and $\chi_{B}$ by a normalized overlap of process matrices \cite{Horodecki99,Bongioanni10},
\begin{equation}
F=\frac{\mathrm{Tr}[\chi\chi_{B}]}{\mathrm{Tr}[\chi]\mathrm{Tr}[\chi_{B}]}.
\end{equation}
This quantum process fidelity satisfies $0\leq F\leq 1$ and if $F=1$, then the implemented operation $\hat{W}$ is a purity-preserving quantum filter which 
coincides with the target operation $\hat{B}$ up to a constant prefactor, $\hat{W}=\sqrt{P_I}\hat{B}$.
The quantum process fidelity $F$ is plotted in Fig.~7(a), where the red dots represent experimental data and the solid line indicates prediction of a theoretical model
of the experimental setup, which accounts for imperfect two photon interference with visibility $\mathcal{V}=0.94$ and imperfections of the partially polarizing beam splitter PPBS,
whose measured transmittances $T_V=0.687$ and $T_H=0.981$ slightly differ from the nominal transmittances $T_V=\frac{2}{3}$ and $T_H=1$.  
This model is similar to the model presented in the Appendix of Ref. \cite{Micuda15}, where we refer the reader for more details. We can see in Fig.~7(a) that the theoretical model correctly predicts the 
qualitative dependence of fidelity on $T_0$, but the quantitative agreement with the experiment is not exact. This remaining discrepancy is likely 
caused by other effects that may reduce the fidelity, such as phase fluctuations and imperfections of the various wave plates and other optical components.

\begin{figure}[!t!]
\centering
\includegraphics[width=0.99\linewidth]{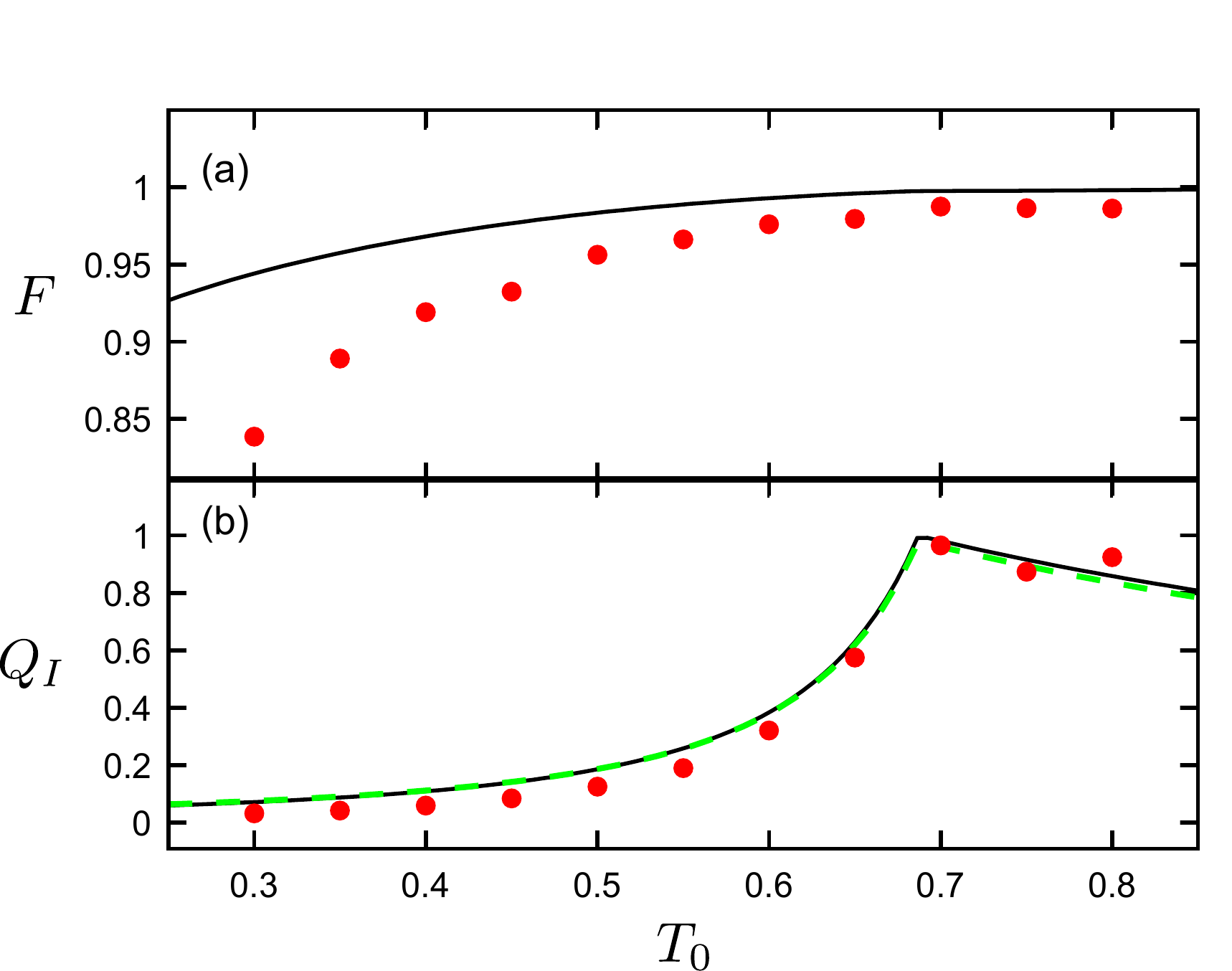}
\caption{(Color online) (a) Fidelity $F$ of the implemented operation $\chi$ with the ideal operation $\hat{B}$, and (b) implementation probability quantified by parameter $Q_I$, are plotted in dependence on the target transmittance $T_0$. 
Red dots depict experimental data and the black curves represent the prediction of a theoretical model. The green dashed curve in panel (b) is the implementation probability $P_I$
of an ideal protocol as discussed in Sec. III of the paper. Note that the solid and dashed lines in panel (b) almost coincide. The statistical error bars are smaller than size of the symbols.}
 \label{fig:overview}
\end{figure}

The matrices $\chi_{B}$ representing the ideal operations $\hat{B}$ are real and their imaginary parts exactly vanish. 
Due to various experimental imperfections, we observe small nonzero imaginary parts of the experimentally determined matrices $\chi$, which
 increase with decreasing $T_0$. It can be seen from Fig.~6 that the dominant imaginary components correspond to a residual phase shift of state $|\bm{11}\rangle$, which cannot be compensated 
 by local single-qubit unitary transformations on qubits A and B.

If the actually implemented operation reads $\hat{W}=\sqrt{P_I}\hat{B}$, then the implementation probability $P_I$ can be determined as a ratio of traces of  $\chi$ and 
$\chi_{B}$. We can generalize this to imperfect implementations and define a quantity 
\begin{equation}
Q_I=\frac{\mathrm{Tr}[\chi]}{\mathrm{Tr}[\chi_{B}]},
\end{equation}
where $\mathrm{Tr}[\chi_{B}]=2-2T_0+4T_0^2$. Generally, $Q_I$ can be larger than $1$. Nevertheless, in case of a high fidelity between $\chi$ and $\chi_{B}$ the parameter $Q_I$ provides a suitable quantification 
of the implementation probability of the target operation $\hat{B}$.
In order to experimentally determine $Q_I$ we have measured $36$ additional reference coincidences $R_m$, one for each two-qubit input state. These reference coincidences
were recorded with qubits prepared in state $|\bm{00}\rangle$ and all half wave plates set to full transmittance. 
The parameter $Q_I$ was estimated from the experimental data as follows,
\begin{equation}
Q_I=\frac{4}{\mathrm{Tr}[\chi_{B}]} \frac{\bar{C}}{\bar{R}},
\end{equation}
where  the average coincidence rates read
\begin{equation}
\bar{C}=\frac{1}{36^2} \sum_{m=1}^{36} \sum_{n=1}^{36} C_{mn}, \qquad \bar{R}=\frac{1}{36} \sum_{m=1}^{36}R_{m}.
\end{equation}
The experimentally determined implementation probability $Q_I$ is plotted in Fig. 7(b) together with the prediction of our theoretical 
model which accounts for imperfect two-photon interference and imperfections of PPBS. For comparison, the figure 
also shows the implementation probability $P_I$ for perfect error-free realization of the protocol, as given by Eqs. (\ref{PIplusBS}) and (\ref{PIminusBS}).
The two theoretical curves are almost identical and the measured dependence of $Q_I$ on $T_0$ closely follows the theoretical predictions.

\begin{figure}[t]
\centering
\includegraphics[width=0.99\linewidth]{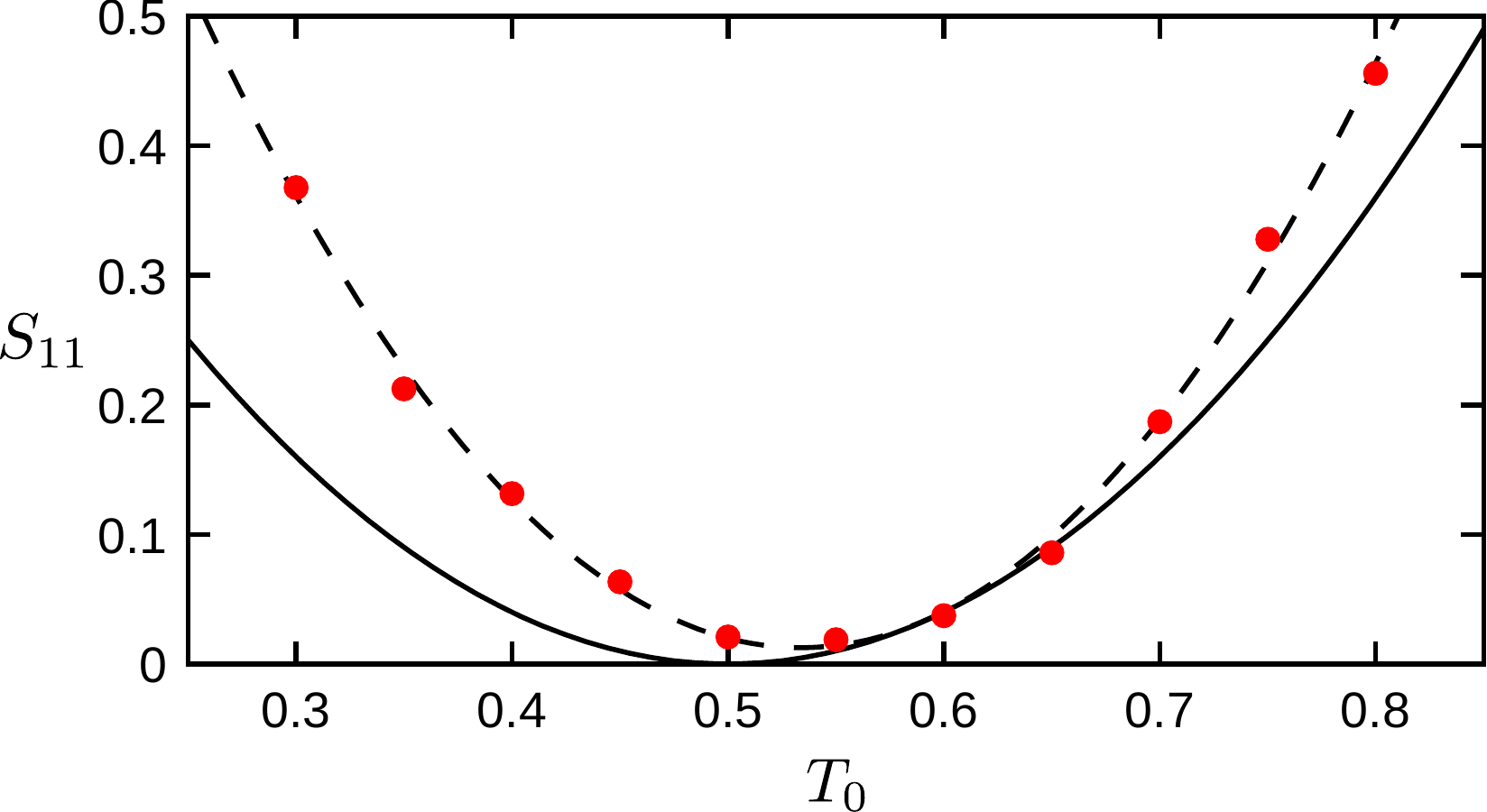}
\caption{(Color online) Observation of the Hong-Ou-Mandel effect. The normalized coincidence rate $S_{11}$ is plotted as a function of the target beam splitter transmittance $T_0$. 
The red dots represent experimental data. The black solid line represents the theoretical curve for an ideal beam splitter and the dashed line 
is the best quadratic fit to the data.}
\label{fig:HOM}
\end{figure}

Bunching of  two photons interfering at a balanced beam splitter is a fundamental nonclassical phenomenon that is exploited 
in countless quantum optics and quantum information processing schemes and experiments \cite{Bachor04,Kok07,Pan12}. 
We can observe the presence of photon bunching and the Hong-Ou-Mandel effect in Fig.~6, where we can see that the matrix element 
\begin{equation}
S_{11}=\langle \bm{1111}|\tilde{\chi}|\bm{1111}\rangle= \frac{\langle \bm{1111}|\chi|\bm{1111}\rangle}{\langle \bm{0000}|\chi|\bm{0000}\rangle}
\label{S11}
\end{equation}
 practically vanishes for $T_0=0.5$. It follows from Eq. (\ref{S11}) that $S_{11}$ can be interpreted as a ratio of probability of coincidence detection of two 
 photons in output modes $A_1$ and $B_1$ when they are injected in modes $A_1$ and $B_1$, and probability of coincidence detection of photons in output modes $A_0$ and $B_0$ when they are injected in modes $A_0$ and $B_0$. 
 For a perfect scheme implementing operation $\hat{W}=\sqrt{P_I}\hat{B}$ we have 
 \begin{equation}
S_{11}=(r_0^2-t_0^2)^2=(1-2T_0)^2,
 \end{equation}
irrespective of the value of the implementation probability $P_I$. The experimentally determined $S_{11}$ is plotted in Fig.~8. 
We can see that the experimental data follow a quadratic dependence on $T_0$ and the minimum is located close to $T_0=\frac{1}{2}$. 
Remarkably, the HOM dip is formed by destructive quantum interference of three alternatives, instead of two as in the ordinary two-photon interference at a beam splitter. Specifically, 
the photons injected in modes $A_1$ and $B_1$ can reach the output modes $A_1$ and $B_1$ as follows: (i) the signal photon in mode $A_1$ is transmitted through BS$_X$ and BS$_Y$ and both photons are transmitted through BS; (ii)
 the signal photon in mode $A_1$ is transmitted through BS$_X$ and BS$_Y$ and both photons are reflected at BS; and (iii) the signal photon avoids the central beam splitter BS by being reflected at both 
BS$_X$ and BS$_Y$ and the idler photon is transmitted through BS to the output mode $B_1$. The data plotted in Fig.~8 thus represent an elementary example of a multi-photon interference in a multiport interferometer, 
a phenomenon that has been recently intensively investigated in the context of boson sampling \cite{Broome13,Spring13,Tillmann13,Crespi13,Carolan14}.

\section{Controlled phase gate}
So far we have considered the interferometric coupling defined by Eq. (\ref{BSstates}). In the present Section we show that the proposed procedure is applicable to a wider class of qubit-qubit interactions.
In particular, we shall consider a deterministic interaction governed by a Hamiltonian $H=\hbar\kappa |11\rangle\langle 11|$. The resulting two-qubit unitary controlled phase gate $\hat{U}(\phi)=e^{-iHt/\hbar}$ 
introduces a phase shift $\phi=\kappa t$ if and only if both qubits are in the state $|1\rangle$,
\begin{equation}
\hat{U}(\phi)= |00\rangle\langle 00| +|01\rangle\langle 01|+ |10\rangle\langle 10| +e^{i\phi}|11\rangle\langle 11|.
\end{equation}
The controlled-Z gate is just a special case of a controlled phase gate, with $\phi=\pi$.
For ease of notation, we use in this Section the ordinary symbols $|0\rangle$ and $|1\rangle$ to denote the computational basis states of a qubit, since there is no risk of confusion with Fock states. 

We assume that the phase shift $\phi$ is fixed and we would like to convert $\hat{U}(\phi)$ to a modified interaction $\hat{U}(\theta)$ with a different phase shift $\theta$.
This can be conditionally accomplished by a scheme similar to that shown in Fig. 2(a). 
 The procedure requires an auxiliary state $|2\rangle_A$ of particle A, to which the qubit state $|1\rangle_A$ is coupled before the inter-qubit interaction as follows,
\begin{equation}
|1\rangle_A \rightarrow t_X|1\rangle_A+r_X|2\rangle_A, \qquad |2\rangle_A\rightarrow t_X|2\rangle_A-r_X|1\rangle_A.
\end{equation}
Here $t_X$ and $r_X$ are generally complex coefficients satisfying $|t_X|^2+|r_X|^2=1$. After interaction between the qubits, states 
$|1\rangle_A$ and $|2\rangle_A$ are coupled again, this time with coupling parameters  $t_Y$, $r_Y$. 
Particle A is then projected onto the qubit subspace and the amplitude of state $|0\rangle_A$ is attenuated according to $|0\rangle_A \rightarrow t_A|0\rangle_A$.
In Ref. \cite{Micuda15} we have shown that this procedure allows to conditionally implement a CZ gate between the qubits for any $0< \phi\leq \pi$. Here we extend this concept and explicitly show that 
the resulting conditional phase shift $\theta$ can be arbitrarily tuned. Note that, in contrast to the case of the interferometric coupling considered in the previous sections, here the quantum filtering needs 
to be applied only to the particle A.

\begin{figure}[t]
\centering
\includegraphics[width=\linewidth]{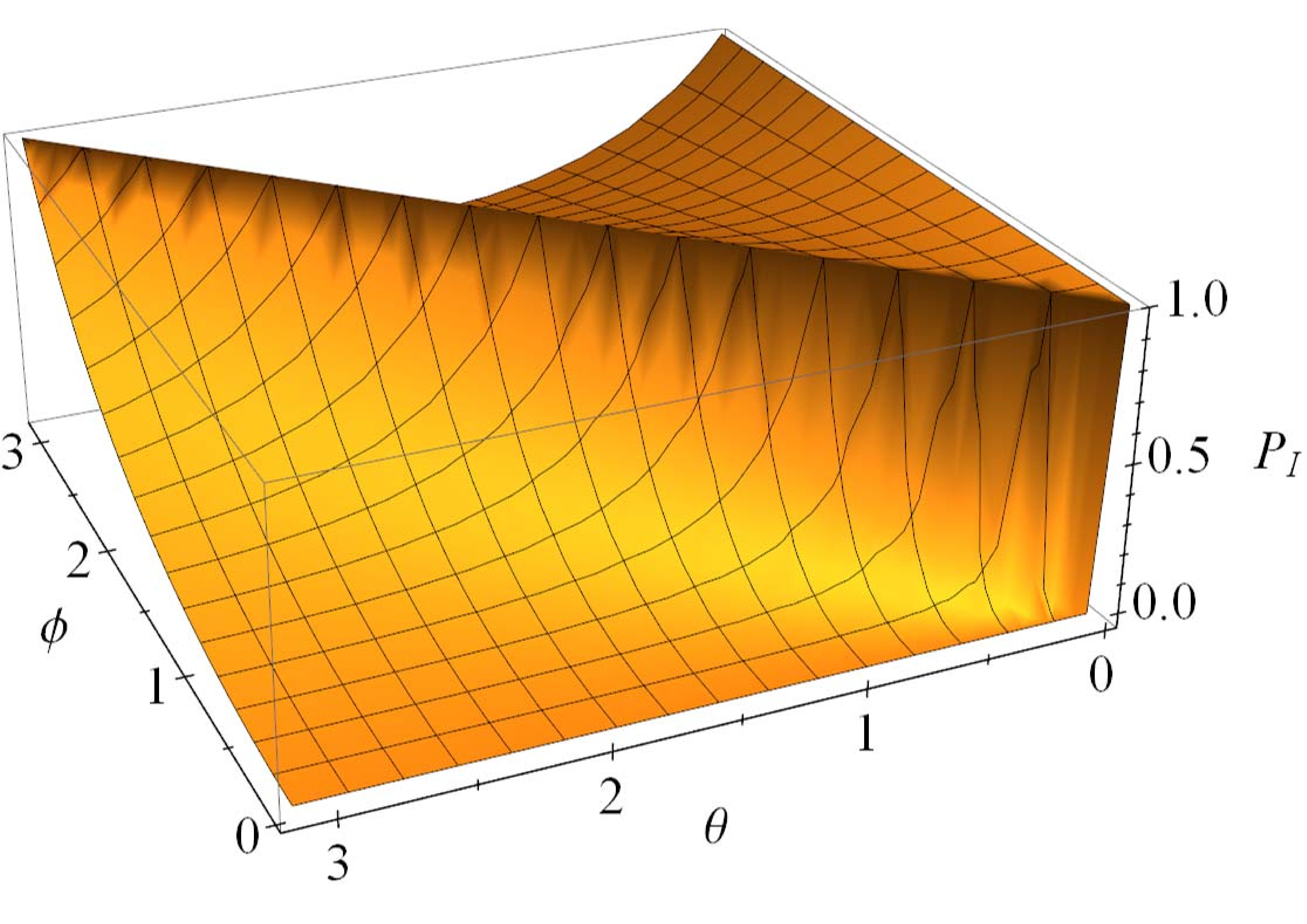}
\caption{(Color online) Dependence of the implementation probability $P_I$ of a two-qubit controlled phase gate $\hat{U}(\theta)$ on the target phase shift $\theta$ and the actual phase shift $\phi$.}
\end{figure}

After some algebra similar to that reported in Sec. IIA we find that the controlled phase gate $\hat{U}(\theta)$ is conditionally implemented provided that the following conditions are satisfied,
\begin{equation}
t_A=t_X t_Y-r_Xr_Y,
\end{equation}
and
\begin{equation}
\frac{r_X r_Y}{t_X t_Y}=\frac{e^{i\phi}-e^{i\theta}}{1-e^{i\theta}}.
\label{trCphase}
\end{equation}
The implementation probability $P_I=|t_A|^2$ is maximized when 
\begin{equation}
|t_X|^2=|t_Y|^2=\frac{\left|\sin \frac{\theta}{2}\right|}{\left|\sin \frac{\theta}{2}\right|+\left|\sin \frac{\theta-\phi}{2}\right|},
\end{equation}
and the phases of the amplitude transmittances and reflectances  should be chosen such that  Eq. (\ref{trCphase}) holds.
For this optimal configuration, we get 
\begin{equation}
P_I= \left (\frac{\left|\sin \frac{\phi}{2}\right|}{\left|\sin \frac{\theta}{2}\right|+\left|\sin \frac{\theta-\phi}{2}\right|} \right)^2.
\end{equation}
The implementation probability $P_I$ is plotted in Fig.~9 in dependence on $\phi$ and $\theta$. Note that $P_I$ is nonzero for any $0<\phi < \pi$, hence the above  procedure enables complete conditional 
tuning and control of the effective conditional phase shift $\theta$.

\section{Conclusions}
In summary, we have investigated in detail the ability to conditionally control and enhance interaction between two qubits by
coupling one or both particles carrying the qubits to an auxiliary quantum state or mode.
We have seen that quantum filtering is an essential part of our procedure, and its probabilistic nature is the price to pay for the enhancement of the interaction. 
The method is not limited to the 
interferometric coupling embodied in our work by two-photon interference at a beam splitter. As we have illustrated in the final part of our paper, 
the proposed concept of conditional interaction enhancement can be applied also to other qubit-qubit interactions such as that resulting in a controlled phase gate between the qubits. 
We have utilized the linear optics platform as a suitable test bed for demonstration and verification of the feasibility and robustness of the proposed scheme, which is mainly intended for configurations where the interaction 
is limited e.g. due to inherently small coupling strength, or due to  noise or decoherence that puts a limit on the total achievable interaction time. 
We hope that our scheme may find applications for instance in heterogeneous quantum networks \cite{Kimble08} or in quantum optomechanics where photons are coupled to phononic excitations of a mechanical oscillator \cite{optbook}.

\begin{acknowledgments}
This work was supported by the Czech Science Foundation (GA13-20319S). 
R.S. acknowledges support by Palacky University (IGA-PrF-2015-005).
\end{acknowledgments}

\end{document}